\documentclass[12pt]{iopart}

\usepackage{iopams,cite,graphicx}
\begin{document}

\title[]{Nanoscale chemical mapping of laser-solubilized silk}

\author{Meguya Ryu$^1$, Hanae Kobayashi$^2$, Armandas Bal\v{c}ytis$^{3,4}$, Xuewen Wang$^3$, Jitraporn Vongsvivut$^5$, Jingliang Li$^6$,\\\protect Norio Urayama$^2$, Vygantas Mizeikis$^7$, Mark Tobin$^5$,\\\protect Saulius Juodkazis$^{3,8}$, Junko Morikawa$^1$ }
\address{$^1$Tokyo Institute of Technology, Meguro-ku, Tokyo
152-8550, Japan}
\address{$^2$Nihon Thermal Consulting Co., 1-5-11, Nishi-Sinjuku,
Shinjuku-ku, Tokyo 160-0023, Japan}
\address{$^3$Nanotechnology facility, Swinburne University of
Technology, John st., Hawthorn, 3122 VIC, Australia}
\address{$^4$Center for Physical Sciences and Technology,
Savanoriu ave. 231, LT-02300 Vilnius, Lithuania}
\address{$^5$Infrared Microspectroscopy Beamline, Australian
Synchrotron, Clayton, VIC 3168, Australia}
\address{$^6$Institute for Frontier Materials, Deakin University,
Geelong, VIC 3220 Australia}
\address{$^7$Research Institute of Electronics, Shizuoka
University, Naka Ku, 3-5-3-1 Johoku, Hamamatsu, Shizuoka 4328561,
Japan}
\address{$^8$Melbourne Center for Nanofabrication, Australian
National Fabrication Facility, Clayton~3168, Melbourne, Australia}
\ead{sjuodkazis@swin.edu.au; morikawa.j.aa@m.titech.ac.jp}
\vspace{10pt}
\begin{indented}
\item[25] August 2017
\end{indented}

\begin{abstract}
A water soluble amorphous form of silk was made by ultra-short
laser pulse irradiation and detected by nanoscale IR mapping. An
optical absorption-induced nanoscale surface expansion was probed
to yield the spectral response of silk at IR molecular
fingerprinting wavelengths with a high $\sim 20$~nm spatial
resolution defined by the tip of the probe. Silk microtomed
sections of 1-5~$\mu$m in thickness were prepared for nanoscale
spectroscopy and a laser was used to induce amorphisation.
Comparison of silk absorbance measurements carried out by
table-top and synchrotron Fourier transform IR spectroscopy proved
that chemical imaging obtained at high spatial resolution and
specificity (able to discriminate between amorphous and
crystalline silk) is reliably achieved by nanoscale IR. A
nanoscale material characterization using synchrotron IR radiation
is discussed.
\end{abstract}

%

%
%
%

\section{Introduction}

In analytical material science, absorption of IR light is used for
fingerprinting (chemical imaging) of particular molecules,
specific compounds, and provides insight into interactions in
their immediate vicinity. However, challenges arise when this
information needs to be obtained from sub-wavelength and
sub-cellular dimensions, in particular at IR and terahertz
spectral bands of absorption~\cite{Dazzi1} or
scattering~\cite{Hillenbrand}. Optical properties of
sub-wavelength structures and patterns have opened an entirely new
direction in photonics and design of highly efficient optical
elements for control of intensity, phase, polarisation, spin and
orbital momenta of light based on flat planar geometries, yet rich
in nanoscale features~\cite{Kuznetsov,Kruk}. We aim at reaching
that level of control in the domain of chemical spectral imaging.
There, interpretation of data from near-field requires further
knowledge of probe interaction with substrate, phase information
of the reflected/transmitted light from sub-wavelength structures
to reveal complex peculiarities of light-matter interactions at
nanoscale and is now advancing with strongly concentrated
efforts~\cite{Woessner,Ni,Huber,Zenin,Khanikaev,Fei}. Recently, an
electron tunneling control by a single-cycle terahertz pulse
illuminated onto a tip of a scanning transmission microscope (STM)
needle was demonstrated at 10~V/nm fields~\cite{Katayama}. STM
reaches an atomic precision in surface probing and its
spectroscopic characterisation and can be carried out on a water
surface~\cite{Guo}.

\begin{figure}
\begin{center}
\includegraphics[width=9.5cm]{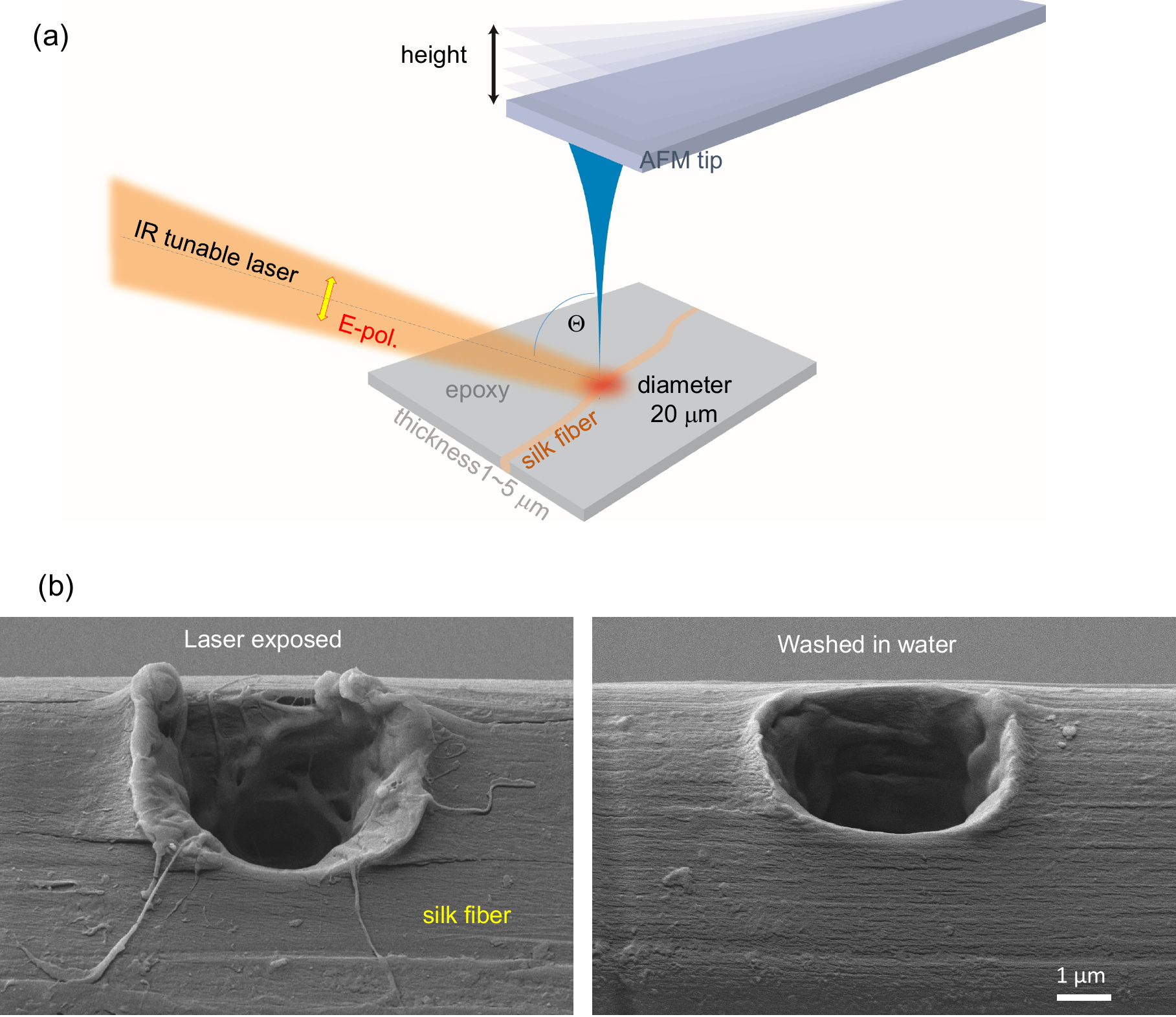} %
\caption{Principle of nano-IR: an AFM tip follows height changes
induced by a spectral sweep of excitation at IR wavelengths ($\sim
6~\mu$m). IR laser beam is p-polarised and incident at
$\Theta\simeq 60^\circ$ angle to minimise reflective losses (close
to the Brewster angle) and to illuminate a much larger sample area
while AFM readout occurs from a AFM needle contact ($\sim 20$~nm)
continuously scanned with 0.1~Hz frequency and digitised to obtain
map with $x\times y = 25\times 100$~nm$^2$ pixels. AFM and nano-IR
mapping was carried out at a selected excitation wavelength. (b)
SEM images of silk fiber melted/amorphised by a single laser pulse
as-fabricated and after immersion into water; laser wavelength
515~nm, pulse duration 230~fs, focused with objective lens with
numerical aperture $NA = 0.5$, pulse energy 425~nJ, linear
polarisation was along the fiber (different fibers exposed at the
same conditions were used for the two images).} \label{f-princ}
\end{center}
\end{figure}

Absorbance spectra quantify and identify the resonant molecules,
chemical structures through their individual or collective
excitations as detected in transmission (or inferred from
reflection). By sweeping excitation wavelength through the finger
printing region of a particular material, the usual optical
excitation relaxation pathway ending with a thermal energy
deposition into host material can be sensitively measured using an
atomic force microscopy (AFM) approach~\cite{Dazzi,Carminati}
which opened up a rapidly growing AFM-IR field~\cite{NComm15}
(also known as nano-IR; Fig.~\ref{f-princ}(a)). How reliably one
can determine IR properties with an AFM nano-tip based on the
thermal expansion is currently still under debate due to a lack of
knowledge of the actual anisotropy of thermal and mechanical
properties at the nanoscale, 3D molecular conformation, alignment,
and the interaction volume. To establish the correspondence
between nano-IR and spectroscopy it is necessary to compare IR
spectral imaging in near-field and far-field modes with nano-IR.
Another challenge of the nano-IR technique is that a very thin
sub-micrometer film has to be prepared and mounted on a thermally
conductive substrate. Microtomed thin sections usually have
thicknesses above the optimum and need to be embedded into an
epoxy host which interferes with the nano-IR signal from
micro-specimens.

In order to demonstrate that microtomed slices of samples with
features at or below micrometers in lateral cross-section can be
measured using nano-IR and provide reliable IR spectral
information we acquired absorbance spectra from areas down to
single pixel hyper-spectral resolution accessible on table-top
FT-IR spectrometers, and at high-resolution which was achieved
with a solid immersion lens using high-brightness synchrotron
FT-IR microscopy. Such comparison of IR properties read out from
nanoscale and far-field was carried out in this study using silk -
a bio-polymer complex in its structure comprised of crystalline
and amorphous building
blocks~\cite{Yoshioka,Hu,Yazawa16,17sr7419}.

Here, micrometer-thick slices of silk were used to measure thermal
expansion under a specific wavelength of excitation with $\sim
20$-nm-sharp AFM tips and to compare with high-resolution spectra
measured with ATR FT-IR method using synchrotron radiation at the
Infrared Microspectroscopy (IRM) Beamline (Australian Synchrotron)
as well as a table-top FT-IR spectrometer. Amorphisation of silk
induced by single ultra-short laser pulses~\cite{17sr7419} has
been spectroscopically recognised using nano-IR.

\begin{figure}[tb]
\begin{center}
\includegraphics[width=8.0cm]{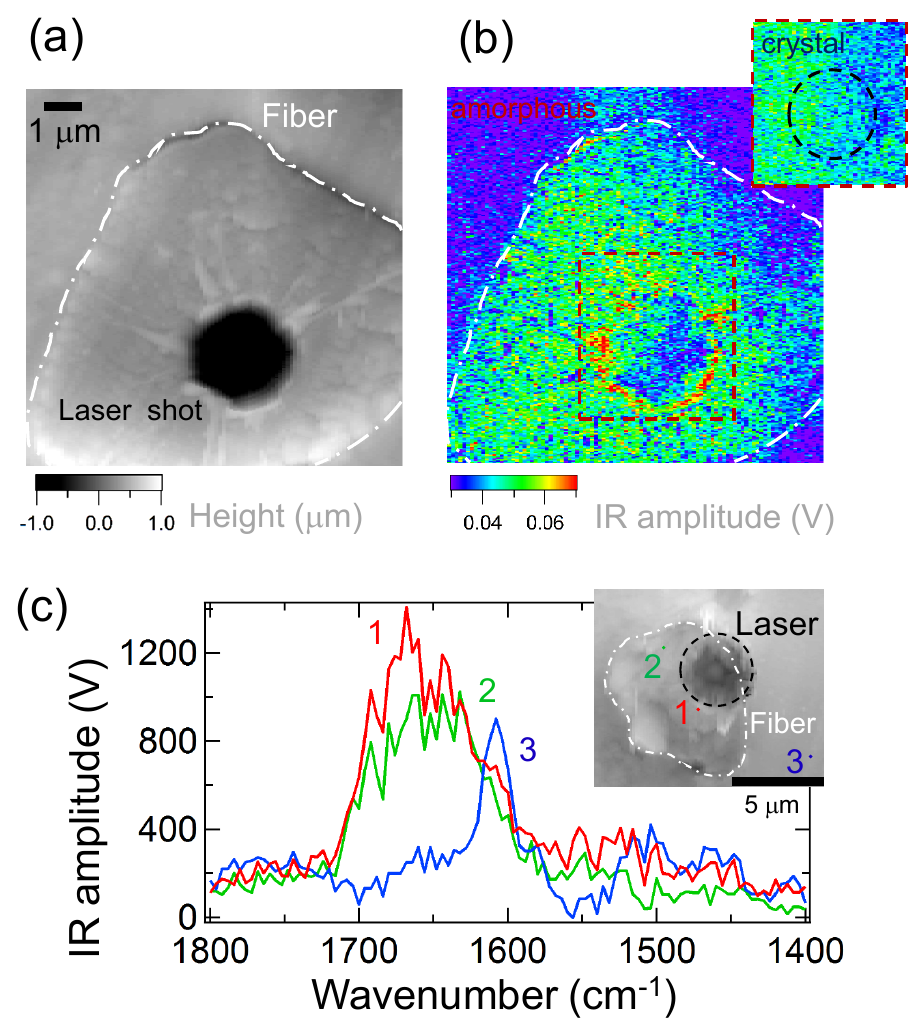}
\caption{AFM (a) and chemical map at Amide I 1660~cm$^{-1}$ band
(b) measured with nano-IR; inset in (b) shows the ablation pit
region imaged at crystalline band at 1700~cm$^{-1}$. (c) Point
spectra measured on epoxy, amorphous-rich, and predominantly
crystalline regions of the T-cross-section of a silk fiber; inset
shows AFM image of a microtome slice of a silk fiber in the epoxy
matrix and laser irradiated crater region.} \label{f-spec}
\end{center}
\end{figure}

\section{Samples and methods}

Domestic (\emph{Bombyx mori}) silk fibers, stripped of their
sericin rich cladding~\cite{16b054101}, were probed during the
spectroscopic imaging experiments. Pulsed laser radiation was used
to induce local structural modifications of silk. The AFM-based
nano-IR experiments with $\sim$25-nm-diameter tips were
benchmarked against more conventional methods, such as attenuated
total reflectance (ATR) at the Australian Synchrotron IRM Beamline
($\sim 1.9~\mu$m resolution), and a table-top FT-IR spectrometer
($\sim 6~\mu$m resolution).

For this set of experiments method-agnostic silk samples, in the
form of thin slices, had to be prepared. For cross-sectional
observation, the natural silk fibers were aligned and embedded
into an epoxy adhesive (jER 828, Mitsubishi Chemical Co., Ltd.).
Fibers fixed in the epoxy matrix were microtomed into
1-5~$\mu$m-thick slices which are mechanically robust enough to be
measured using standard FT-IR setups without any supporting
substrate. This was particularly important to increase sensitivity
of the far-field absorbance measurements and to diminish
reflective losses. Both longitudinal (L) and transverse (T) slices
of the silk fibers were prepared by microtome (RV-240, Yamato
Khoki Industrial Co., Ltd). The slices were thinner than the
original silk fibers. For synchrotron ATR FT-IR, an aluminum disk
was used to support the thin silk sections when they were brought
into contact with a 100~$\mu$m diameter facet tip of the Ge ATR
hemisphere (refractive index $n = 4$).

Synchrotron ATR-FTIR mapping measurement was performed using a
in-house developed ATR-FTIR device at the IR Microspectroscopy
Beamline, which has a high-speed, high-resolution surface
characterisation capabilities with spatial resolution down to
$~1.9~\mu$m~\cite{Pimm}. A 100~$\mu$m tip ATR accessory for a
FT-IR spectrometer (Hyperion 3000, Bruker) with Ge contact lens of
$NA = n\sin\varphi \simeq 2.4$ of refractive index $n = 4$ and
$\varphi = 36.9^\circ$ a half-angle of the focusing cone was used.
A deep sub-wavelength resolution $r = 0.61\lambda_{IR}/NA \simeq
1.5~\mu$m is achievable for the IR wavelengths of interest at the
Amide band of $\lambda_{IR} = 1600-1700$~cm$^{-1}$ or 6.25 -
5.9~$\mu$m. The nano-IR experiments, i.e., an AFM readout of the
height changes in response to IR sample excitation, were carried
out with nano-IR2 (Anasys Instruments, Santa Barbara, CA) tool
with a $\sim$20~nm diameter tip. During continuous scan over the
selected region with 0.1~Hz frequency, digitisation was carried
out resulting in $25\times 100$~nm$^2$ pixels in $x\times y$. The
oscillation frequency of the AFM needle was 190~kHz. Table-top
FT-IR spectrometer (Spotlight, PerkinElmer) was used with a
detector array with $\sim 6~\mu$m pixel resolution.

Localized modification of silk was carried out via exposure to
515~nm wavelength and 230~fs duration pulses (Pharos, Light
Conversion Ltd.) in an integrated industrial laser fabrication
setup (Workshop of Photonics, Ltd.). Fibers were imaged and laser
radiation was focused using an objective lens of numerical
aperture $NA = 0.5$ (Mitutoyo). Single pulse exposures were
carried out with pulse energy, $E_p$. Optical and electron
scanning microscopy (SEM) were used for structural
characterisation of the laser modified regions.

\begin{figure}[tb]
\begin{center}
\includegraphics[width=9.5cm]{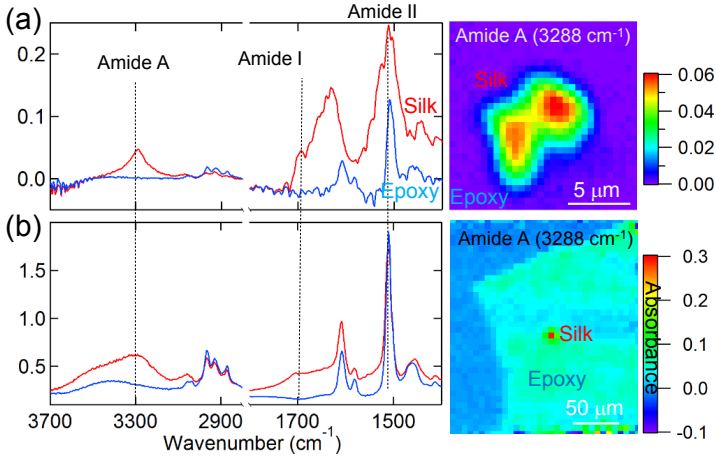}
\caption{Point spectra and chemical maps of T cross sections of
silk (red) in epoxy (blue). (a) Synchrotron ATR-FTIR spectra and
its chemical map based on Amide A band. (b) Table-top FT-IR
spectra and the corresponding Amide A chemical map. Note that the
CCD array has the pixel size of $6.25\times 6.25~\mu$m$^2$
comparable with the T-cross-section of the silk fiber. Chemical
maps (on the right-side) are presented as measured (without
smoothing).} \label{f-supp}
\end{center}
\end{figure}

\section{Results and Discussion}

A highly crystalline natural silk fiber can be thermally
amorphised only through rapid $2\times 10^3$~K/s thermal quenching
from melt~\cite{Cebe} as demonstrated for a very tiny amounts of
silk measured in nanograms (1~ng occupies a sphere of 5.7~$\mu$m
diameter). Amorphous silk is water soluble and can be used as 3D
printing material for scaffolds, desorbable
implants~\cite{Hotz,Li,Tao}, bio-resists~\cite{Sun}, and even be
utilised as an electron beam resist~\cite{Kim,15a11863}.
Amorphous-to-crystalline transition of silk fibroin is usually
achieved via a simple water and alcohol bath processing at
moderately elevated temperatures $\sim 80^\circ$~\cite{15a11863}.
An UV 266~nm wavelength nanosecond pulsed laser irradiation was
also used to enrich amorphous silk fibroin with crystalline
$\beta$-sheets~\cite{Tsuboi1}.

To realise a fast thermal quenching and to retrieve the amorphous
silk phase~\cite{17sr7419}, we used ultrashort 230~fs duration and
515~nm wavelength laser pulses tightly focused into focal spot of
$d = 1.22\lambda/NA \simeq 1.3~\mu$m; the numerical aperture of
the objective lens was $NA = 0.5$ and the wavelength was $\lambda
= 515$~nm. Single pulse irradiation was carried out to ablate
nanograms of silk and create a molten phase which is thermally
quenched fast enough~\cite{Cebe} to prevent crystallisation
(Fig.~\ref{f-princ}(b)). Threshold of optically recognisable
modification of silk fiber during laser irradiation was at 8~nJ
which corresponds to 2.8~TW/cm$^2$ average power (0.6~J/cm$^2$
fluence) per pulse and is typical for polymer
glasses~\cite{16le16133}.

Nano-IR spectrum, the low-frequency band of the AFM detected
height changes in response to a spectral sweep of excitation at
the IR absorbance bands at 1~kHz frequency was measured on a
transverse (T) microtome cross section of silk fiber embedded into
a micro-thin epoxy (Fig.~\ref{f-spec}(a)). Single wavelength
chemical map measured at the specific Amide I band of
1660~cm$^{-1}$ associated with the amorphous silk components is
shown in (b) with clearly discernable amorphous rim of the
ablation crater, which, in contrast, is not recognisable at the
crystalline $\beta$-sheet region of 1700~cm$^{-1}$ (inset of (b)).
Typical single point measurement spectra are shown in
Fig.~\ref{f-spec}(c) for the Amide I region with clear distinction
between epoxy matrix, amorphous, and crystalline counterparts of
silk. Amorphous silk of the molten quenched phase has only 200~nm
thickness as observed by SEM and AFM, however, it was
distinguished using illumination at the corresponding absorption
bands ($\lambda_{ex} \simeq 6~\mu$m).

Thin microtomed slices placed on a high thermal conductivity
substrate have been shown to enhance speed of nano-IR imaging
since a thermalisation time scale is $t_{th}\simeq \rho c_p
h^2/\eta$~\cite{Katzenmeyer}, where $h$ is the thickness of the
film, $\eta$ is thermal conductivity, $\rho$ is the mass density,
and $c_p$ is the specific heat capacity.

\section{Conclusions and outlook}

It was shown that by using micro-thin slices which are
sub-wavelength at the IR spectral range of interest, it was
possible to obtain spectral band readout using nano-IR method -
the surface height changes due to thermal expansion following the
absorbance spectrum of silk. Amorphous silk created by ultra-fast
thermal quenching at the irradiation location of fs-laser pulse
was distinguished on the chemical map and imaged with lateral
resolution defined by digitisation $x\times y \equiv 25\times
100$~nm$^2$ and can potentially reach the limit defined by
tip~\cite{Wang2017} which was $20\times 20$~nm$^2$ in this study.
Chemical mapping result acquired using the nano-IR method is
consistent with far-field spectroscopy of silk carried out with
table-top and synchrotron FT-IR (see Fig.~\ref{f-supp}(b)). The
far-field FT-IR detectors have a typical pixel size of $\sim
6~\mu$m which limits the resolution to the entire T-cross-section
of the silk fiber, however, a good correspondence with nanoscale
IR spectral mapping is confirmed between different spectroscopic
methods. Laser-induced amorphisation of crystalline silk has been
spectroscopically resolved with high spatial resolution.

We can envisage, that a simple principle of the nano-IR technique
allows for a coupling with synchrotron light sources and to
complement it with a phase and amplitude mapping using scanning
near-field microscopy. The high brightness of synchrotron
radiation would also enable a fast mapping required for temporally
resolved evolution of photo or thermally excited processes, and
even a molecular alignment mapping could be realised by using the
four polarisation method~\cite{Hikima}. These functionalities will
bring new developments into a cutting edge nanoscale molecular
characterisation.

\section*{Acknowledgments}
\small{J.M. acknowledges a partial support by a JSPS KAKENHI Grant
No.16K06768. We acknowledge the Swinburne's startup grant for
Nanotechnology facility and partial support via ARC Discovery
DP130101205 and DP170100131 grants. Experiments were carried out
via beamtime project No.~11119 at the Australian Synchrotron IRM
Beamline. Window-on-Photonics R\&D, Ltd. is acknowledged for joint
development grant and laser fabrication facility.}

\vspace{1cm}
\bibliographystyle{iopart-num}
\providecommand{\newblock}{}

\end{document}